\documentclass[10pt,conference]{IEEEtran}
\usepackage{booktabs}
\usepackage{array}
\usepackage{pst-node}
\usepackage{lipsum}
\usepackage{auto-pst-pdf}
\usepackage{epstopdf}
\usepackage{graphicx}
\usepackage{amsmath}
\usepackage{amssymb}
\usepackage{xcolor}
\usepackage{xfrac}
\usepackage{braket}
\usepackage{subcaption}
\usepackage{comment}
\usepackage{float}
\usepackage{mathtools}
\usepackage{tikz}
\usetikzlibrary{quantikz2}
\usepackage[]{mdframed}
\usepackage{multirow}
\usepackage{algorithm}
\usepackage{algpseudocode}
\usepackage{soul}
\usepackage{hyperref}

\definecolor{green}{rgb}{0.2, 0.9, 0.0}

\newcommand{\abs}[1]{\lvert #1 \rvert}

\newcommand{\kett}[1]{| #1\rangle}
\newcommand{\braa}[1]{\langle #1 |}

\newcommand{\expec}[3]{\langle #1\lvert #2 \rvert #3\rangle}

\DeclareMathOperator*{\argmin}{arg\,min}

\begin{document}
%

\title{Towards secondary structure prediction of longer mRNA sequences using a quantum-centric optimization scheme}

\author{\IEEEauthorblockN{Vaibhaw Kumar}
\IEEEauthorblockA{\textit{IBM Quantum}\\
New York, USA \\
vaibhaw.kumar@ibm.com}
\and
\IEEEauthorblockN{Dimitris Alevras}
\IEEEauthorblockA{\textit{IBM Quantum}\\
New York, USA \\
alevras@us.ibm.com}
\and
\IEEEauthorblockN{Mihir Metkar}
\IEEEauthorblockA{\textit{Moderna}\\
Cambridge, USA \\
mihir.metkar@modernatx.com}
\and
\IEEEauthorblockN{Eline Welling}
\IEEEauthorblockA{\textit{Fermioniq B.V.}\\
Amsterdam, NL \\
eline@fermioniq.com}
\and
\IEEEauthorblockN{Chris Cade}
\IEEEauthorblockA{\textit{Fermioniq B.V.}\\
Amsterdam, NL \\
chris@fermioniq.com}
\and
\IEEEauthorblockN{Ido Niesen}
\IEEEauthorblockA{\textit{Fermioniq B.V.}\\
Amsterdam, NL \\
ido@fermioniq.com}
\and
\IEEEauthorblockN{Triet Friedhoff}
\IEEEauthorblockA{\textit{IBM Quantum}\\
New York, USA \\
triet.nguyen-beck@ibm.com}
\and

\IEEEauthorblockN{Jae-Eun Park}
\IEEEauthorblockA{\textit{IBM Quantum}\\
New York, USA \\
parkje@us.ibm.com}
\and
\IEEEauthorblockN{Saurabh Shivpuje}
\IEEEauthorblockA{\textit{Moderna}\\
Cambridge, USA \\
sshivpuje@modernatx.com}
\and
\IEEEauthorblockN{Mariana LaDue}
\IEEEauthorblockA{\textit{IBM Quantum}\\
New York, USA \\
mariana.ladue@ibm.com}
\and
\IEEEauthorblockN{\qquad\qquad\qquad\qquad\qquad}
\and
\IEEEauthorblockN{Wade Davis}
\IEEEauthorblockA{\textit{Moderna}\\
Cambridge, USA \\
wade.davis@modernatx.com}
\and
\IEEEauthorblockN{Alexey Galda}
\IEEEauthorblockA{\textit{Moderna}\\
Cambridge, USA \\
alexey.galda@modernatx.com}
\and
\IEEEauthorblockN{\qquad\qquad\qquad\qquad\qquad}
}
\bstctlcite{IEEEexample:BSTcontrol}
\maketitle

\begin{abstract}

Accurate prediction of mRNA secondary structure is critical for understanding gene expression, translation efficiency, and advancing mRNA-based therapeutics. However, the combinatorial complexity of possible foldings, especially in long sequences, poses significant computational challenges for classical algorithms. In this work, we propose a scalable, quantum-centric optimization framework that integrates quantum sampling with classical post-processing to tackle this problem. Building on a Quadratic Unconstrained Binary Optimization (QUBO) formulation of the mRNA folding task, we develop two complementary workflows: a Conditional Value at Risk (CVaR)-based variational quantum algorithm enhanced with gauge transformations and local search, and an Instantaneous Quantum Polynomial (IQP) circuit-based scheme where training is done classically and sampling is delegated to quantum hardware. We demonstrate the effectiveness of these approaches using IBM quantum processors, solving problem instances with up to 156 qubits and circuits containing up to 950 nonlocal gates, corresponding to mRNA sequences of up to 60 nucleotides. Additionally, we validate scalability of the CVaR algorithm on a tensor network simulator, reaching up to 354 qubits in noiseless settings. These results demonstrate the growing practical capabilities of hybrid quantum-classical methods for tackling large-scale biological optimization problems.

\end{abstract}

\IEEEpeerreviewmaketitle

\section{Introduction}\label{sec:introduction}

Messenger RNA (mRNA) serves as a fundamental component in gene expression, acting as the intermediary between genomic DNA and the cellular machinery responsible for protein synthesis. Its biological function is intricately linked to its secondary structure, which arises from specific base-pairing interactions and folding patterns~\cite{vicens2022thoughts}. This structural aspect is vital for understanding gene expression, regulation, translation, and the dynamics of mRNA degradation~\cite{metkar2024tailor}, and is therefore central to both basic molecular biology and the development of mRNA-based therapeutics~\cite{qin2022mrna}.

mRNA secondary structure prediction represents a substantial computational challenge due to the myriad potential folding configurations and associated energy considerations. The structural complexity of RNA is amplified by its six backbone torsion angles – compared to only two in proteins – increasing the range of possible structures~\cite{zhang2024predicting}. This complexity is reflected in the Protein Data Bank, where RNA structures, particularly mRNAs, are significantly underrepresented, highlighting the need for precise computational prediction tools. Moreover, predicting structures with pseudoknots is classified as an NP-complete problem~\cite{lyngso2000rna}. Traditional methods such as the Zuker algorithm and its derivatives, MFold and ViennaRNA, use dynamic programming to evaluate thermodynamic stability~\cite{zuker198920, zuker2003mfold, schuster1997rna}. Enhancements such as stochastic context-free grammars in tools like Infernal~\cite{nawrocki2013infernal} aid in predicting conserved structures, and the integration of machine learning in ContextFold enhances accuracy by utilizing both local and global contextual information~\cite{zakov2011rich}. Nevertheless, the inherent limitations of classical methods, particularly their difficulties in managing the combinatorial complexity of possible structures, have prompted a growing interest in utilizing quantum computing to address these challenges.

In a recent study~\cite{alevras2024mrna}, we evaluated the feasibility of using quantum computers to predict mRNA secondary structure, using up to 80 qubits on IBM Eagle and next-generation Heron processors. These experiments, which involved mRNA sequences of up to 42 nucleotides, represented the largest instance of such problems tackled by quantum computing to date. However, classical algorithms~\cite{yang2025advances} can process mRNA sequences with hundreds or thousands of nucleotides, but only by relying on strong approximations, such as excluding pseudoknots. Scalable alternatives are needed to model these structures more accurately. The development of quantum computing approaches offers a promising path forward. This work demonstrates substantial progress in that direction by applying quantum-centric workflows to increasingly complex instances, thereby laying a robust groundwork for further breakthroughs in the field.

There has been a significant push towards developing quantum capabilities beyond utility scale, which enables tackling increasingly complex optimization problems. However, a majority of utility-scale studies on gate-based quantum computers~\cite{barron2024provable,miessen2024benchmarking,sack2024large,sachdeva2024quantum,romero2024bias,nodar2024scaling,cadavid2024bias,maciejewski2024improving,leclerc2025implementing}, focus on problems that align closely with the architecture of available quantum hardware. This is primarily due to the significant overhead incurred in circuit depth and gate counts when the problem structure is incompatible with the hardware layout. 

Variational Quantum Algorithms (VQAs) offer a versatile approach to quantum optimization, that can be run on hardware-efficient circuits, making them applicable to a broad range of problems. Despite this adaptability, VQAs face significant challenges during the optimization phase. The task of fine-tuning circuit parameters to minimize the cost function is inherently NP-hard, often complicated by the presence of multiple local minima that can trap the optimization process~\cite{bittel2021training,fontana2022non,anschuetz2022quantum, larocca2022diagnosing}. Moreover, a prevalent issue in VQAs is the emergence of barren plateaus~\cite{mcclean2018barren,larocca2024review}. Although the challenges of multiple local minima and barren plateaus are formidable, they do not entirely preclude the potential discovery of effective parameter trajectories that may bypass these hurdles in practical settings. Further research suggests that circuits that successfully navigate these issues may, on average, be amenable to classical simulation~\cite{cerezo2023does}. Encouragingly, this insight provides an avenue to develop pragmatic approaches in which circuits could be trained classically, and quantum computers could then be employed primarily for sampling from these optimized circuits. Furthermore, the expressivity of the circuits significantly affects the performance of VQAs~\cite{tikku2022circuit}, underscoring the need for a balanced approach in circuit design that pays attention to the class of circuits that are difficult to classically simulate while ensuring trainability, and expressivity.

As we enter the utility era of quantum computing, the capability of quantum computing to address problems beyond the reach of conventional brute-force classical methods is rapidly becoming a reality. This advancement necessitates a careful reevaluation of how quantum and classical resources are distributed and integrated, especially as we address increasingly complex challenges beyond the utility scale. The development of a quantum-centric approach, which strategically combines quantum processors with classical compute nodes, has shown promising results in the realm of chemistry~\cite{robledo2024chemistry, kaliakin2024accurate}. This model, where the quantum processor is the focal point of computation complemented by classical resources, provides a robust blueprint for tackling large-scale problems.

In this paper, we propose two quantum-centric approaches that orchestrate the assignment of workloads between quantum and classical compute nodes, allowing us to tackle relatively large problems. Both strategies use variational circuits. The first approach builds upon a previous work~\cite{alevras2024mrna}, where the optimal circuit parameters are found using a feedback loop between the quantum processor and classical nodes, as it is commonly done for a VQA. Following Ref.~\cite{barkoutsos2020improving}, parameter updates are driven by the Conditional Value at Risk (CVaR) as the objective function, based on the samples collected from the quantum processor. In this work, we additionally introduce a gauge transformation on the original Hamiltonian as a means to mitigate noise in hardware samples to improve convergence. We also implement a parameter threshold to further boost convergence during circuit executions on quantum hardware. To increase the likelihood of reaching the ground state, the mitigated samples undergo additional post-processing through a shallow local search on classical nodes. This multi-tiered approach leverages the unique strengths of both quantum and classical computing resources to ensure optimal convergence of the VQA scheme to the lowest-energy solution.

To investigate the scalability of the CVaR VQA approach, we employ tensor network methods~\cite{schollwock2011density, lykov2022tensor, orus2014practical} to simulate the quantum circuits of the VQA. By decomposing the wave function into a network of smaller tensors, each with a bond dimension that captures the amount of entanglement present in the quantum state, tensor network simulations can scale to much larger systems than is possible with state-vector simulators. States with limited entanglement can be described exactly by a tensor network state with a modest bond dimension. For highly entangled states, the bond dimension required to represent the state exactly grows exponentially with the number of qubits and must be artificially restricted -- to ensure that the tensor network state fits in classical memory -- at the cost of an approximation error that increases with the degree of entanglement~\cite{zhou2020,ayral2023}. Despite this limitation, our tensor network simulations have proven crucial in validating the effectiveness of the CVaR VQA approach for tackling large problem instances, surpassing the capabilities of other types of classical simulators and of currently available quantum hardware. By exploiting the linear topology and relatively shallow depth of the VQA circuits, we successfully solve mRNA secondary structure prediction problems with up to 354 qubits. Although this validation is limited to the ideal case in which the quantum circuits are noiseless, it provides essential confirmation that the CVaR VQA scheme could remain viable on future quantum hardware, and that it can potentially scale successfully to hundreds of qubits.

In our second approach, we introduce an optimization scheme that utilizes parameterized Instantaneous Quantum Polynomial (IQP) circuits~\cite{jordan2009permutational, shepherd2010quantum}. We specifically select IQP circuits  because the expectation values of a state prepared by such a circuit can be efficiently estimated using classical computers. However, sampling from the output distributions of an IQP circuit is generally considered classically difficult~\cite{bremner2011classical,bremner2016average,marshall2024improved}. It is worth noting that the presence of noise can complicate these hardness assumptions~\cite{Rajakumar_2025}. We outline a strategy, also described in Refs.~\cite{recio2025iqpopt, recio2025train}, wherein the training of a parameterized IQP circuit occurs entirely on a classical computer, using expectation values as the objective metric. The quantum computer's role is then limited to sampling bitstrings from the circuit containing optimal parameters. To enhance the quality of these samples, we employ an error mitigation technique that utilizes classically computed single-qubit expectation values to adjust for noise present in the quantum hardware outputs. Similar to our first method, we also apply a lightweight local search algorithm to further enhance the probability of reaching the ground state. This dual-stage optimization harnesses the strengths of both classical and quantum computing, providing a robust framework to improve convergence.

Our results provide a promising path towards solving longer mRNA sequences using a quantum-centric approach. Using this workflow, we have successfully examined mRNA sequences of length up to 60 nucleotides corresponding to problem sizes that require up to 156 qubits. The experiments involve quantum circuits with up to 950  gates and circuit depths up to 112. This scale of quantum computation not only highlights the growing capabilities of quantum hardware but also underscores the potential for significant advancements in understanding complex biological structures through quantum-assisted methodologies.

The paper is organized as follows. In Section~\ref{sec:QUBO}, we provide a concise description of the Quadratic Unconstrained Binary Optimization (QUBO) formulation of the mRNA secondary structure prediction problem. Section~\ref{sec:methods} is dedicated to detailing the methodologies employed in this research, as well as discussing the results obtained from the examination of several mRNA sequences. Lastly, Section~\ref{sec:conclusions} offers a summary of our findings and provides an outlook on future directions of this research.

\begin{figure*}
 \begin{subfigure}[t]{.47\linewidth}
    \includegraphics[width=\linewidth]{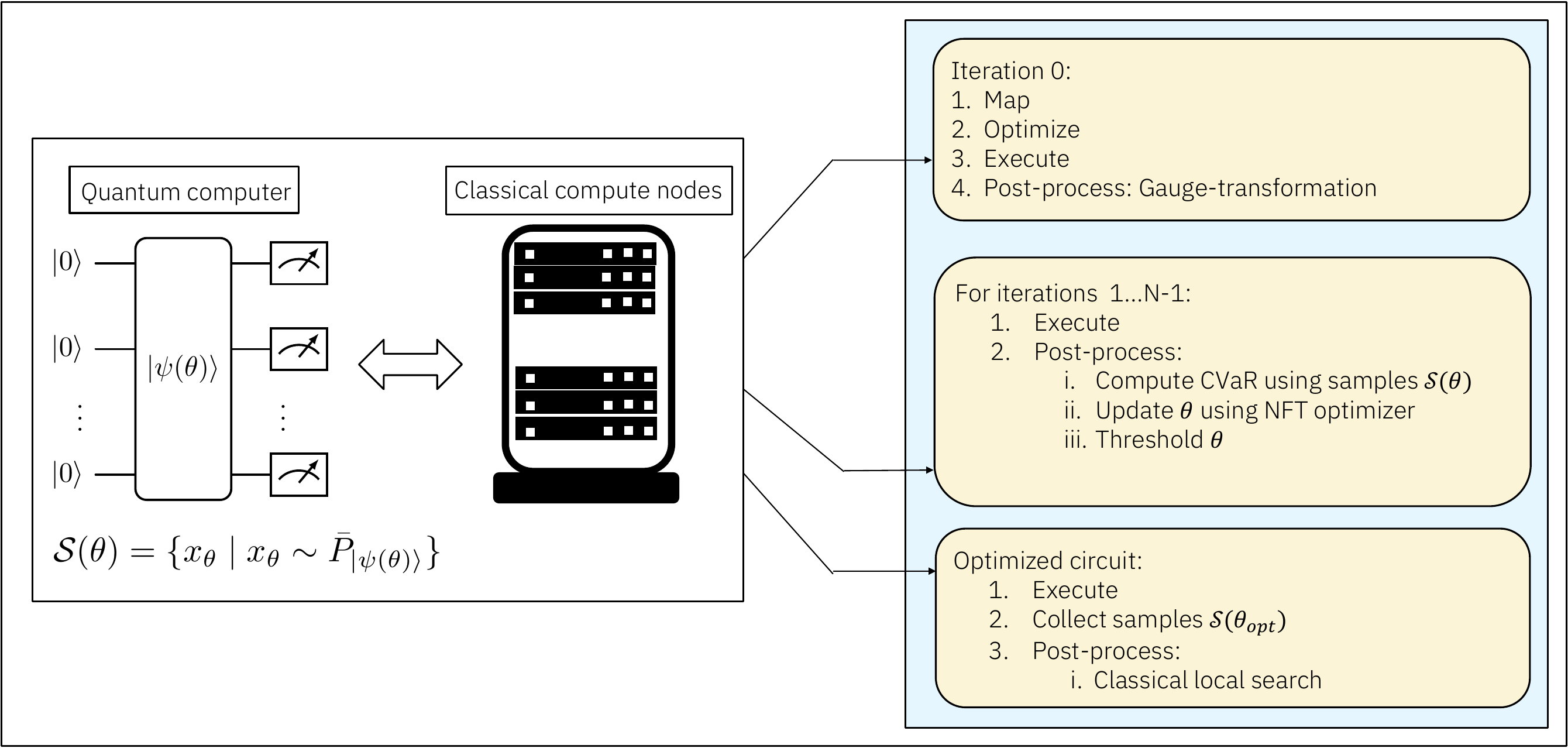}%
  \caption{}
  \label{fig: sampling schematic}
  \end{subfigure}\hfill                 
\begin{subfigure}[t]{0.5\linewidth}
  \includegraphics[width=\linewidth]{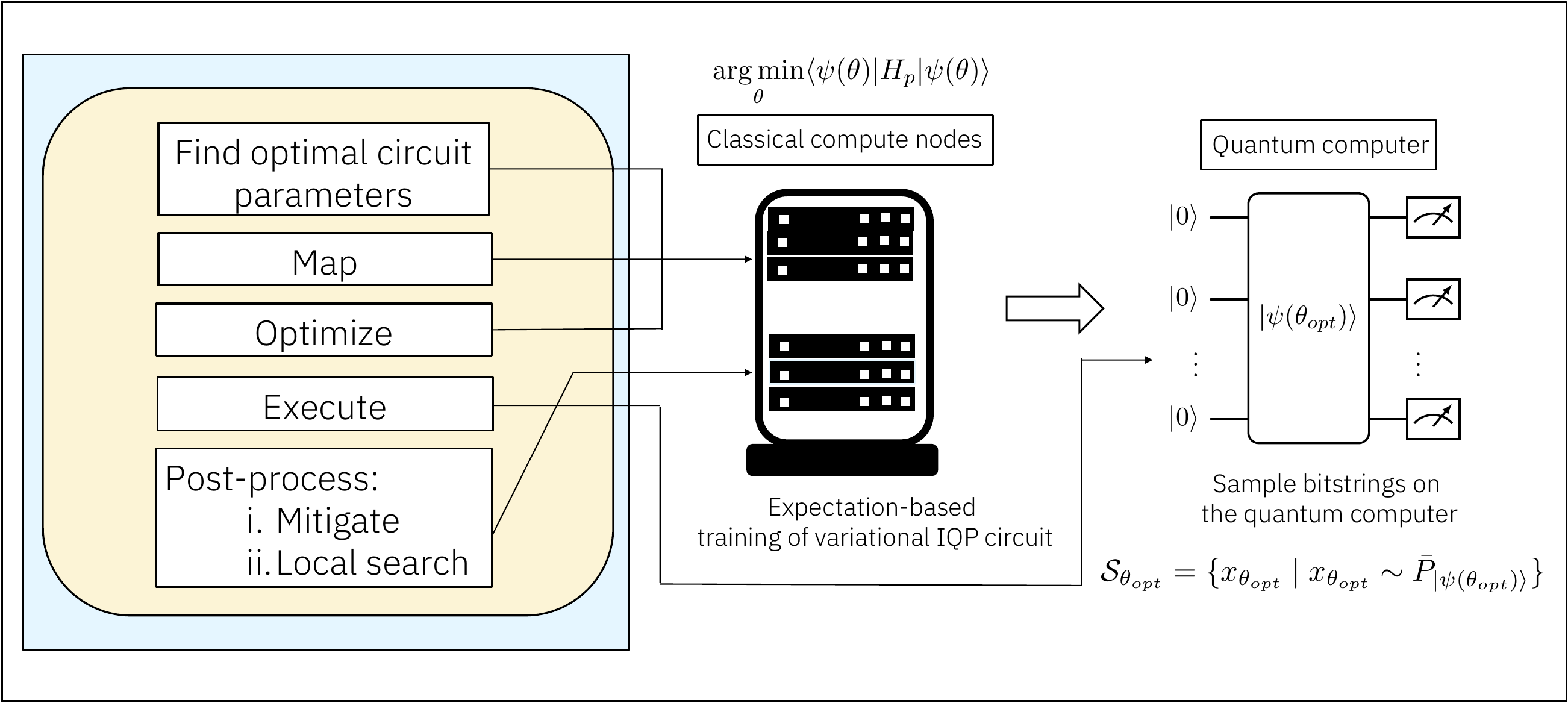}
  \caption{}
  \label{fig: iqp schematic}
\end{subfigure}
\caption{a) Schematic for the quantum-centric architecture for sampling-based quantum optimization using CVaR. The training of the variational circuit is carried out by collecting hardware samples $S_{\theta}$ characterized by the noisy distribution $\Bar{P}_{\psi(\theta)}$ and processing them on the classical compute nodes. b) Schematic of the approach based on the IQP circuit. Training of the variational circuit is conducted only on the classical compute nodes. The quantum processor is used for sampling from the optimized circuit.}
\end{figure*}

\section{QUBO formalism}\label{sec:QUBO}
In this Section, we provide an overview of the QUBO formulation used to model the mRNA secondary structure prediction problem. 

Each position in an RNA sequence, indexed by $i = 1,\ldots,n$, is occupied by one of the nucleotide bases: $U,A,C,$ or $G$. Base pairing between two positions $i$ and $j$ is deemed valid if the bases form one of the following pairs: $\{(AU), (UA), (CG), (GC), (GU), (UG)\}$. Further, the decision variable represents a quartet which is defined as two consecutive base pairs, often referred to as stacked pairs which is represented by the tuple $(i, j, i+1, j-1)$, where $(i, j)$ and $(i + 1, j - 1)$ indicate the positions of the two consecutive base pairs. 

Following the framework established in Ref.~\cite{gusfield2019integer}, we define the quartet (stacked pair) variable as::

$$x(i,j,i+1,j-1)=\left \{
\begin{array}{rl}
1 &\text{if stacking occurs between base}\\
& \text{pairs $(i,j)$ and $(i+1,j-1)$} \\
0 &\text{otherwise.}
\end{array}
\right .
$$ 
Note that these variables are established exclusively for valid base pairs. The model incorporates constraints based on specific rules to ensure valid RNA structures. One crucial rule is that each base can be paired with only one other base, effectively preventing multiple pairings for any single base within the sequence. Additionally, two quartets cannot be selected simultaneously if their base pairs cross each other, which is formalized in the constraint:
\begin{equation} \label{eq:mis}
x(i,j,i+1,j-1) + x(k,\ell,k+1,\ell-1) \le 1.   
\end{equation}
This type of constraint is also observed in the independent set problem~\cite{glover2018tutorial}. Defining 
$q_i$ as the variable for the $i^{th}$ quartet and denoting $Q$ as the set of all valid quartets, $QC$ as the set of crossing quartet pairs, $QS(q_i)$ as the set of quartets that can be stacked with $q_i$, and $\mathit{QUA}$ as the set of stacked quartets ending in a $(UA)$ pair, we express the corresponding QUBO problem as follows:
\begin{align}\label{eq:QUBO} 
     \min &\sum_{q_i \in Q} e_{q_i} q_i  +
r\sum_{q_i \in Q}{\sum_{q_j \in QS(q_i)}}q_i q_j \nonumber \\  
& + p \sum_{q_i \in Q} \sum_{q_j \in \mathit{QUA}} q_i (1-q_j) + t \sum_{q_i,q_j \in QC} q_i q_j. 
\end{align}

The QUBO formulation in Eq.~(\ref{eq:QUBO}) includes terms that capture the energy contributions of individual quartets within the RNA structure. Each quartet variable $q_i$ is assigned a free energy $e_{q_i}$ derived from empirical thermodynamic data, as described in Ref.~\cite{turner-mathews}, and contributes to the objective function through the term $e_{q_i} q_i$. Stability contributions from consecutive quartets, which are known to enhance RNA structural stability, are explicitly rewarded in the model. This is captured by the reward factor $r$ and is incorporated into the objective function as $r\sum_{q_i \in Q}{\sum_{q_j \in QS(q_i)}}q_i q_j$. In addition, the model imposes penalties to discourage structurally unfavorable configurations. For example, sequences ending in a less stable $(UA)$ pair are penalized using a factor $p$, implemented through the term $p\sum_{q_i \in Q} \sum_{q_j \in \mathit{QUA}} q_i (1-q_j)$, following the energetics described in Ref.~\cite{turner-mathews}. The constraints that prevent the formation of crossing quartets are modeled with a penalty factor $t$ as $t \sum_{q_i,q_j \in QC} q_i q_j$. For computational implementation, we utilize the Qiskit~\cite{Qiskit} framework that facilitates the conversion of such problems into QUBO format and assists in calculating penalties for the constraints. The transformation from QUBO's Boolean variables to spin variables, which are then represented using Pauli Z matrices, is essential for formulating the Hamiltonian
   $H_p = \sum_i Z_i + \sum_i \sum_{j(i)}J_{ij} Z_i Z_{j(i)},$ \,
where $j(i)$ denotes all variables that share a quadratic term with variable $x_i$ in the original QUBO. 

The QUBO problems in this work, derived from the secondary structure prediction of mRNA sequences, typically exhibit high edge densities in their corresponding problem graphs, ranging from approximately 0.7 to 0.85. This indicates a high level of connectivity between variables, reflecting the complex interactions within mRNA structures.

\section{Methods and Results}\label{sec:methods}
\subsection{Sampling-based quantum optimization using CVaR}
\label{sampling-based}
In Fig.~\ref{fig: sampling schematic}, we present the quantum-centric workflow for sampling-based quantum optimization. This methodology uses a variational circuit characterized by a two-local ansatz illustrated in Fig.~\ref{fig: two-local ansatz} to prepare the parameterized state $\kett{\psi(\theta)}$. The search for the ground state is executed collaboratively between quantum processors and classical nodes, which are closely integrated within a quantum-centric compute cluster to maximize computational efficiency.

In the first iteration, both the mapping and optimization steps are performed on the classical node. The two-qubit CZ gates of the two-layer ($p=2$) ansatz are arranged in a “pairwise” pattern: in the first layer, qubit $i$ is entangled with qubit $(i+1)$, for all even values of $i$, and in the second layer, qubit $i$ is entangled with qubit $(i+1)$ for all odd values of $i$. These entangling gates are mapped onto the native heavy-hex topology of the IBM quantum hardware. In the optimization step, the circuit is transpiled using Qiskit Primitive V2 transpiler with the optimization level set to 3. Dynamical decoupling is employed for error suppression using the default `XX' sequence. The resulting transpiled circuit is then executed on the quantum processor using the Qiskit's sampling primitive, with $N_{shots}=2^{15}$ measurement samples collected. 

\subsubsection{Gauge-transformed Hamiltonian}
In the pre-fault-tolerance era of quantum computing, mitigating hardware noise is crucial for achieving reliable and accurate results. To address this challenge, we adopt gauge transformations, a method previously explored for reducing the impact of noise in quantum computations~\cite{pelofske2019optimizing,barbosa2021optimizing}. Notably, this approach has also been employed in an adaptive strategy known as Noise Directed Adaptive Remapping (NDAR)~\cite{maciejewski2024improving}, which dynamically applies gauge transformations at each iteration under the assumption that quantum hardware is biased towards an all-zero quantum state.

Bit-flip gauge transformations, in this context, can be interpreted as applying permutations to the diagonal problem Hamiltonian, $H_p \to \bigotimes_{i=1}^n X^{m_i} H_p\bigotimes_{i=1}^n X^{m_i}$ with $m_i \in \{0,1\}$, indicating which qubits undergo the transformation, as described in Ref.~\cite{maciejewski2024improving}. This operation preserves the eigenvalues of $H_p$ while inducing bit-flips in the corresponding eigenstates of the Hamiltonian. For example, in a three-qubit system with initial eigenstates ${E_0:000, E_1:011, \cdots}$, applying a gauge transformation to the middle qubit would modify the eigenstates to ${E_0:0\boldsymbol{1}0, E_1:0\boldsymbol{0}1, \cdots}$.

The suitability of gauge transformations in our context is further highlighted by the structural similarities between our QUBO problem constraints, Eq.~(\ref{eq:mis}), and those found in the graph-theoretic independent set problem, as discussed in Section~\ref{sec:QUBO}. This problem involves selecting a subset of nodes such that no two connected by an edge are included simultaneously. Such a task becomes particularly challenging in dense graphs. For instance, in a complete graph where each node connects to every other node, an all-zero bitstring becomes the only viable solution. Given the high density typical of our QUBO problem graphs, the feasible and optimal solutions are likely to be bitstrings with relatively low Hamming weights. However, if noisy hardware samples generally display Hamming weights higher than those expected from a noiseless circuit, the optimization process can be significantly slowed down. By applying the gauge transformation, we align the characteristics of noisy hardware samples with the relevant eigenstates of the transformed problem Hamiltonian, thus facilitating more rapid and effective convergence during optimization.

\begin{figure*}
 \begin{subfigure}[b]{.25\linewidth}
    \includegraphics[width=\linewidth, trim={5cm 15cm 6.0cm 4cm},clip]{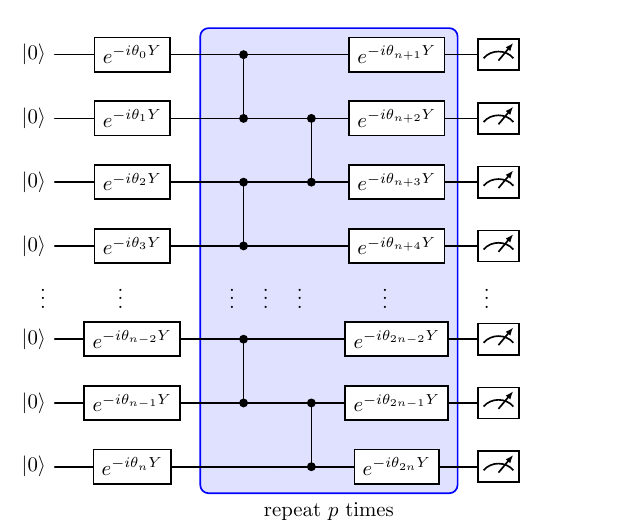}%
    \caption{}
  \label{fig: two-local ansatz}
  \end{subfigure}\hfill              
\begin{subfigure}[b]{0.30\linewidth}
  \includegraphics[width=\linewidth, trim={1.5cm 1.5cm 1.5cm 1.5cm},clip]{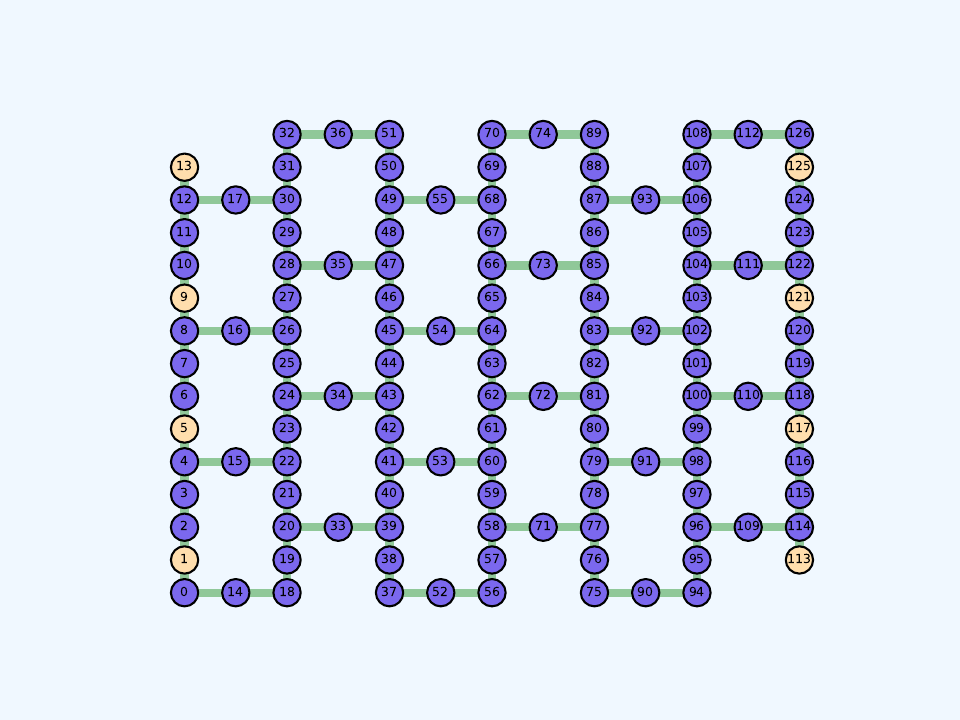}
  \caption{}
   \label{fig: iqp layout}
\end{subfigure} \hfill
\begin{subfigure}[b]{0.33\linewidth}
\includegraphics[width=\linewidth, trim={4.0cm 17.5cm 6.0cm 4.5cm},clip]{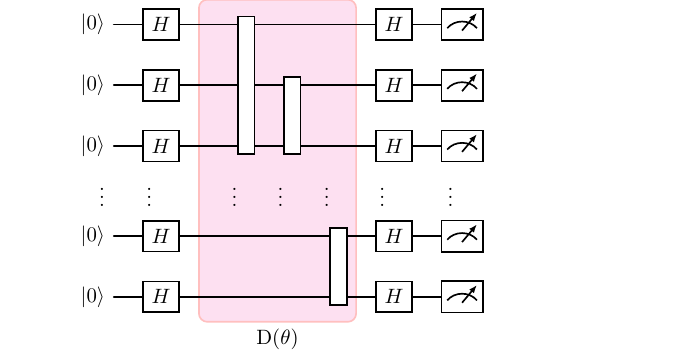}
\caption{}
\label{fig: iqp ansatz}
\end{subfigure}
\caption{a) The two-local ansatz with $p$ repetitions of  characterized by parameterized Pauli-Y single-qubit rotations and two-qubit CZ gates b) IQP-circuit's heavy-hex-based tubular arrangement of two qubit $e^{-i\theta ZZ}$ rotation gates on the \texttt{ibm\_kyiv} heavy-hex layout with edges between yellow-colored qubits along with all the edges on the heavy-hex graph, shown in green c) parameterized IQP-circuit where $D(\theta)$ represents the diagonal unitary of the form $e^{-i\bigotimes_j \theta_j Z_j^{m_j}}$ where $m_j \in \{0, 1\}$.}
\label{fig:iqp scheme}
\end{figure*}

Careful initialization of the circuit parameters allows us to compare noisy hardware samples against samples expected from a noiseless circuit without requiring additional calibration circuits. At iteration 0, we set the rotation angle $\theta_i$ to $\frac{\pi}{4}$ for a small randomly selected fraction $f$ of qubits in the first layer, as shown in Fig.~\ref{fig: two-local ansatz}, while setting other rotation angles to zero, and maintaining the CZ gates unchanged. The application of the gate $e^{-i\frac{\pi}{4} Y}$, which can induce bit flips by creating an equal superposition of $\ket{0}$ and $\ket{1}$, is limited to a few qubits, rest of the qubits stay as $\kett{0}$. As a result, this initialization circuit aims to generate an initial state that predominantly supports bitstrings with relatively low Hamming weight under noiseless settings.

We observe that noisy hardware samples typically exhibit a higher average Hamming weight compared to the ideal, noiseless samples. For instance, a 127-qubit initialization circuit on \texttt{ibm\_kyiv} results in an average Hamming weight of 17, with a standard deviation of 3 when averaged over $2^{15}$ samples, when a Hamming weight of 9 is expected from the noiseless circuit. This discrepancy is likely due to gate and readout errors, causing some qubits that should ideally be in the $\ket{0}$ state to be erroneously read as $\ket{1}$. To address this issue, we compute the noisy single-qubit expectation value $\langle Z_i \rangle_{noisy}$ for all qubits. We identify qubits whose noiseless expectation $\langle Z_i \rangle_{noiseless}$ should be +1 (indicative of the qubit's noiseless state as $\ket{0}$) but display $\langle Z_i \rangle_{noisy} < -p_{th}$, i.e. less than a certain cut-off  $p_{th}$, indicating such qubits are predominantly in the noisy $\ket{1}$ state in the hardware samples. We set the cut-off to $p_{th}=0.8$. A gauge transformation is then applied to these specific qubits in the Hamiltonian. Given the assumption that the optimal bitstring contains only a few bits as ones (owing to low Hamming weight), it is highly probable that this gauge transformation will flip the bit values from $0 \to 1$ at the identified qubit locations in the original optimal bitstring. Consequently, this procedure constructs a new Hamiltonian, calibrated to reflect the hardware sample's Hamming weight distribution more accurately. This transformation is applied only once at iteration 0.

Following the initialization at iteration 0, as illustrated in Fig.~\ref{fig: sampling schematic}, we proceed with further iterations up to the total $N_{iter}$. These subsequent iterations involve repeated cycles between the quantum hardware and classical nodes. After each execution of the circuit, the samples are used on the classical nodes to compute the CVaR value~\cite{barkoutsos2020improving}. CVaR is the mean of the lowest $\alpha$-tail of the energy distribution of the samples, which quantifies the expected value of the best outcomes. This metric has been identified as more robust against noise when used as an objective function in quantum computations~\cite{barron2024provable}. CVaR-based parameter updates are carried out using the classical NFT optimizer~\cite{nakanishi2020sequential} named after the authors Nakanishi, Fujii, and Todo, which has demonstrated enhanced convergence properties in a recent study~\cite{oliv2022evaluating}. For CVaR calculations, we select $\alpha = 0.2$ similar to the value used in Ref.~\cite{alevras2024mrna}.

\subsubsection{Parameter threshold}
To improve convergence on quantum hardware, we implement a circuit parameter thresholding scheme that has been used in recent quantum optimization studies~\cite{cadavid2024bias,romero2024bias}. In this scheme, any circuit parameters where $\abs{\theta_i} < \theta_{th}$ are set to zero. For our experiments, we established, after few quick runs, that applying the threshold $\theta_{th} = 0.06$ substantially improves convergence.

The total number of iterations, $N_{iter}$, is chosen to ensure convergence of the CVaR-based objective to a stable minimum. Since each iteration in the NFT optimization scheme updates only a single parameter, $N_{iter}$ must be large enough to complete at least two full passes over all circuit parameters. The parameters are updated in a randomized order within each epoch.

\subsubsection{Post-processing}
\label{sec: post processing}
After $N_{iter}$ iterations, the circuit characterized by $\theta_{opt}$ is expected to generate sample bitstrings close to the ground state. To enhance the likelihood of identifying the optimal solution, we implement a post-processing step using a computationally efficient classical local-search scheme. This approach has been recognized in recent studies as an effective method to refine raw hardware samples~\cite{sachdeva2024quantum, sciorilli2025towards}. 

The local search algorithm, detailed in Algorithm~\ref{alg:ls}, systematically attempts to improve the bitstrings by applying a single bit flip per step. Each of the $n$ bits undergoes one trial flip, executed in a random order determined by a uniformly distributed sequence. This flip is facilitated by the function $\text{Flip}(\mathbf{x},j)$, targeting the $j^{th}$ bit of bitstring 
$\mathbf{x}$. If a flip reduces the energy, the new bitstring is accepted; otherwise, the change is rejected, and the process continues with the next bit flip attempted on the accepted bitstring at the bit index next in order in the random sequence (see Algorithm~\ref{alg:ls}).

Ideally, if all trial flips are accepted, this process could alter the original bitstring by a Hamming distance of $n$. As a result, the decrease in the objective value that can be introduced by the local search is always 
$\le \Delta = 2 (\sum_i h_i + \sum_{(i,j)\in G} J_{ij})$. Hence, for the local search to find the optimal state with high probability, it is important that the energy difference between the objective value of the mitigated sample $E_{s}$ and the optimal solution $E_{g}$ should not exceed $\Delta$, underscoring the need for mitigated sample energies $E_{s}$ to be sufficiently close to energy $E_{g}$ of the optimal solution.
While various advanced local search methods like Tabu search~\cite{glover1998tabu} exist in the classical literature and can effectively solve problems independently, they generally struggle with problems beyond a certain size. In such cases, the goal of our quantum-centric approach is to leverage the quantum processor to handle the computationally intensive tasks, bringing the system closer to the optimal solution. Subsequently, a classical local solver can be employed to fine-tune and determine the precise optimal solution. 

\begin{algorithm}
\caption{Local Search Algorithm}\label{alg:ls}
\begin{algorithmic}
\Require: A bit string $\mathbf{x} \in \mathcal{S}(\theta_{opt})$ of length $n$; objective function $obj$
\Ensure: A bit string with improved objective
\For {j in \text{Permute}(x)}
\State $\mathbf{x^*} = \text{Flip}(\mathbf{x},j)$
\If{$obj(\mathbf{x^*}) < obj(\mathbf{x})$}
    \State $\mathbf{x} \gets \mathbf{x^*}$
\EndIf
\EndFor
\end{algorithmic}
\end{algorithm}

\subsubsection{Hardware results}
We examine 133 and 150 qubit problems originating from mRNA sequences of length 60 and 48, respectively. The 133 qubit problem, originally sized at 256 qubits, was effectively reduced using the strategy outlined in Appendix~\ref{app: reduction}.  The circuits for these utility scale experiments, comprising 133 and 150 qubits, included 700 and 939 non-local gates respectively on the Heron r2 processor \texttt{ibm\_marrakesh}.

Fig.~\ref{fig:raw and post-processed samples} displays the energy distributions from the final iteration alongside those refined via a local-search-based postprocessing technique. The comparison reveals that while the raw hardware samples are nearly optimal, local search further sharpens the distribution, enhancing the proximity to the ground state. As discussed in Ref.~\cite{alevras2024mrna}, these runs require extensive computational efforts. Specifically, for 200 iterations, the process involves approximately $3 \times N_{shots} \times N_{iter}$ circuit executions, totaling about $600 \times 2^{15} \sim 10^7$ executions demonstrating robustness of IBM Qiskit Runtime primitive driven quantum-centric workflow in handling such large-scale experiments.

The ability to simulate quantum circuits using classical methods is a crucial area that requires thorough exploration to advance toward achieving quantum advantage. Tensor network-based methods~\cite{zhou2020} are a leading choice for the simulation of quantum circuits. Using these tools, we can gain enhanced insight into the simulation overheads associated with various circuit layouts and sizes, as well as assess the effectiveness of these heuristic strategies in an ideal noise-free setting. In the following, we detail a tensor network-based approach specifically designed for the two-local ansatz employed in this work.    

\subsection{Tensor network simulations}
In this Section, we describe the tensor network method developed to simulate the CVaR VQA algorithm, which accurately mirrors its implementation on (noiseless) quantum hardware. 

\subsubsection{Tensor network method}
We utilize a tensor network state to represent the quantum state prepared by the parameterized circuit. After each parameter update, we draw samples from this tensor network state to compute the CVaR, which then informs subsequent parameter adjustments.

The parameterized quantum circuit in our study features a one-dimensional connectivity pattern, making it suitable for representation using a Matrix Product State (MPS)~\cite{fannes1992finitely, ostlund1995thermodynamic, vidal2003efficient}. 
Given the circuit’s shallow depth, characterized by only a few layers of gates, an MPS with an adequately large bond dimension can exactly capture the state generated by the circuit.

\begin{figure}
    \centering
    \includegraphics[width=0.6\linewidth]{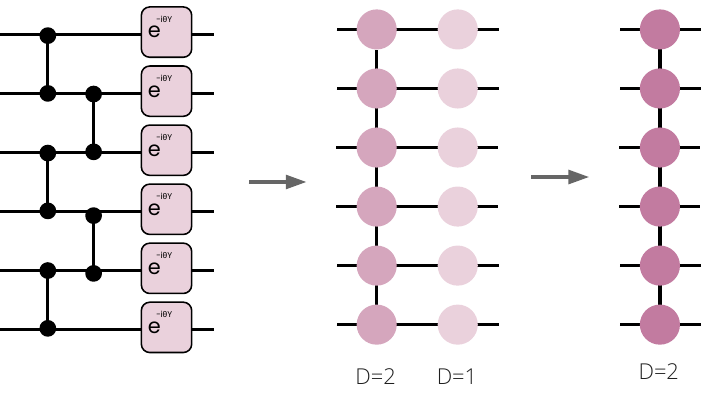}
    \caption{Depiction of a single layer of the parameterized quantum circuit, consisting of an entangling layer of CZ gates followed by a layer of parameterized single-qubit gates. Together, the gates form an $n$-qubit MPO with a bond dimension of 2.}
    \label{fig:CZ_layer_to_mpo}
\end{figure}

The initial state (including the first layer of single-qubit parameterized gates) is a product state and therefore representable by an MPS with a bond dimension of $\chi = 1$. 
The rest of the circuit is composed of layers of CZ gates followed by single-qubit rotations.
A single CZ-gate is a rank-2 tensor, as can be seen from the following decomposition
    $\text{CZ} = \proj{0} \otimes I + \proj{1} \otimes Z \, .$
Therefore, a CZ gate can be decomposed into a two-site Matrix Product Operator (MPO) with a bond dimension of 2. Using the above decomposition, a layer of nearest-neighbor CZ gates on $n$ qubits can be represented by a single $n$-qubit MPO also of bond dimension 2. 
The parameterized single-qubit gates applied after the CZ gates can be contracted into the $n$-qubit CZ MPO without increasing its MPO bond dimension. 
Consequently, each layer of the parameterized circuit can be exactly represented by a single $n$-qubit MPO with an MPO bond dimension of $D=2$, see Figure~\ref{fig:CZ_layer_to_mpo}.

An MPS description of the state prepared by the parameterized circuit is obtained as follows. Starting with the initial product state represented as an MPS of bond dimension $\chi=1$, each layer of the quantum circuit is directly contracted into the MPS. Since each layer is represented by an MPO of bond dimension 2, contracting it exactly into the MPS doubles the MPS bond dimension $\chi$. When a circuit parameter is changed, the MPS can be efficiently updated by only modifying the single local tensor affected by the change. 

After each parameter update, the CVaR is computed by efficiently sampling from the MPS through parallelized sample drawing~\cite{weggemans2024guidable}. The sampling process utilizes the ability to efficiently compute single-site reduced density matrices (RDMs) by combining the MPS with its complex conjugate and contracting across all but one of the physical sites. Sampling then proceeds qubit-by-qubit, where for each qubit the RDM is computed and a sample basis state is drawn from the resulting probability distribution. After drawing a sample, the physical index is fixed and the process moves to the next site. This method ensures that all samples are independent and suitable for CVaR computation.
\begin{figure*}
\begin{subfigure}[t]{0.45\linewidth}
    {\includegraphics[width=0.9\linewidth]{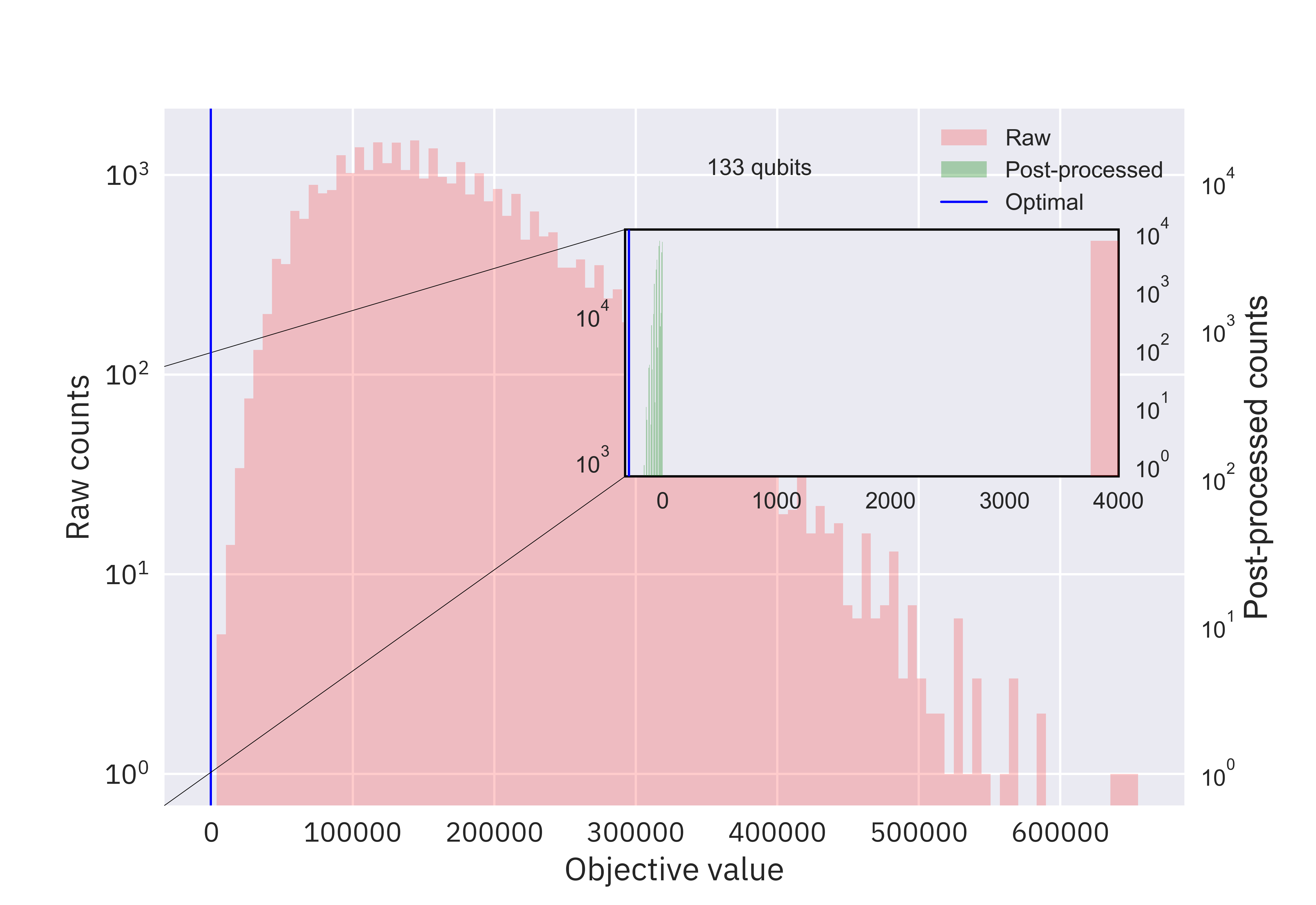}}
\end{subfigure}\hfill
\begin{subfigure}[t]{0.45\linewidth}
\includegraphics[width=0.9\linewidth]{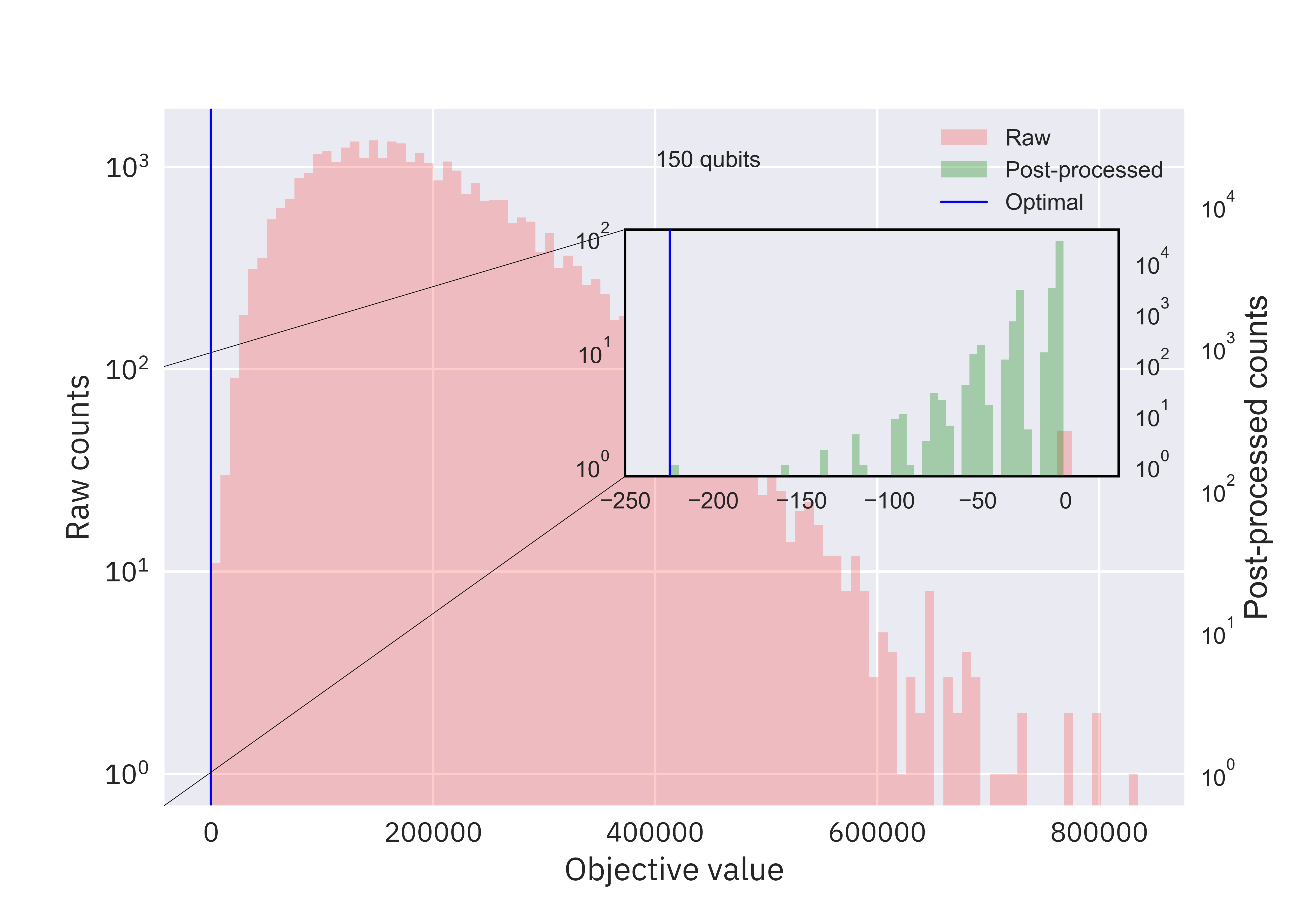}
\end{subfigure}
\caption{The converged unnormalized raw objective value distributions from the hardware experiments 
on \texttt{ibm\_marrakesh} for the a) 133 and b) 150 qubit experiments. Inset shows the post-processed distributions using the local search scheme. Blue solid lines indicate the optimal objective values of -295.4 and -224.65 for the 133 and 150 qubit problems, respectively (found by CPLEX). In both the cases, we are able to find the optimal solution after post-processing.}
\label{fig:raw and post-processed samples}
\end{figure*}

All of the steps above have a runtime that scales (at most) linearly in system size, making the above method of exact circuit contraction well suited for investigating how the algorithm presented in Section~\ref{sampling-based} scales with increasing qubit number.
For example, for a 560-qubit problem instance with two ansatz layers, the total time it took to update a parameter and corresponding local MPS tensor, draw $2^{15}$ samples, and compute the CVaR was less than a second in total (and less than 0.1 second for the $<100$-qubit VQA instances) on a single NVIDIA GH200 GPU.

Increasing the number of layers in the circuit rapidly makes exact contraction computationally intractable. Indeed, a circuit with $p$ layers requires an MPS of bond dimension $\chi = 2^p$ in order to absorb the circuit exactly, and the time it takes to update a local tensor, which only depends on the number of layers, scales as $O(4^p)$. Additionally, the time it takes to draw a sample scales as $O(n\chi^3)$ for the pre-processing plus $O(n \chi^2)$ per sample, resulting in a total runtime of $O(n(8^p + n_s 4^p))$ for $n_s$ samples.
In order to simulate deeper circuits, approximate contraction routines (at the cost of an approximation error), such as those used in IBM's Qiskit MPS simulator, or those used in Fermioniq's \href{https://www.fermioniq.com/ava}{Ava} and in Refs.~\cite{zhou2020, ayral2023, thompson2025} are required. However, these approximate contraction routines are significantly more time consuming, resulting in simulation times reaching minutes (or even hours for particularly challenging circuits) per circuit simulation, making them much less suitable for VQA simulations that easily require thousands of circuit simulations in total.

\subsubsection{Tensor network results}

Using the method described in the previous Section, we explore how the quantum algorithm behaves for larger system sizes. In particular, we simulate the same algorithm discussed in Section~\ref{sampling-based} (using two ansatz layers) for VQA instances of up to 406 qubits to investigate its performance with increasing system size. Table~\ref{mps:table-num-hits} displays how often (out of 100 runs) the algorithm succeeds in finding the optimal solution (hit rate), both with and without classical post-processing. From the table we observe that (i) post-processing increases the hit rate, and (ii) the hit rate decreases rapidly with increasing system size.

\begin{table}[h!]
\centering
\resizebox{\columnwidth}{!}{%
\begin{tabular}{|c||c|c||c|c|}
\hline
\textbf{Qubits} & \multicolumn{2}{|c||}{\textbf{Without post-processing}} & \multicolumn{2}{|c|}{\textbf{With post-processing}} \\
\hline
 & \textbf{Hits} & $\boldsymbol{\gamma}$ & \textbf{Hits} & $\boldsymbol{\gamma}$ \\
\hline\hline
40 & 40 & - & 85 & - \\
80 & 9 & - & 60 & - \\
114 & 2 & - & 12 & - \\
164 & 2 & - & 6 & - \\
212 & 0 & 0.7\% & 1 & - \\
264 & 0 & 4.5\% & 1 & - \\
354 & 0 & 21.4\% & 0 & 10.5\% \\
406 & 0 & 30.3\% & 0 & 20.4\% \\
\hline
\end{tabular}%
}
\caption{Number of times the optimal solution was found out of 100 runs for increasing number of qubits with and without post processing. For systems where the optimal solution was not found, the table shows the relative error to the optimal solution.}
\label{mps:table-num-hits}
\end{table}

When the optimal solution is not found, the relative error, given by $\gamma = \frac{\left| F(\theta_f)_{\text{low}} - F_0 \right|}{|F_0|} \times 100\%$, signifies how close the obtained solution is to the optimal solution, where $F(\theta_f)_{\text{low}}$ represents the lowest value of the objective function (over all samples) obtained at any time during the simulation, and $F_0$ denotes the value of the objective function of the optimal solution found by CPLEX.
From Table~\ref{mps:table-num-hits} we observe that as the system size increases, so does the relative error, indicating that the algorithm has increasing difficulty finding optimal solutions as the number of qubits grows.

These observations raise two key questions: (i) whether running additional optimization epochs improves convergence, and (ii) whether the current ansatz—with only two layers of entangling gates—has sufficient expressivity to produce high-quality solutions as the number of qubits increases. Our simulations indicate that simply adding many more epochs does not lead to significant improvements in the average relative error of the best solution, albeit that the first few epochs added do slightly reduce the error. However, performance can also be modestly improved by resuming optimization from the \textit{best} parameter configuration encountered during an epoch, rather than from the \textit{last} one obtained throughout an epoch. Applying this strategy, we successfully solve the 354-qubit problem instance, correctly finding the known optimal solution.

Increasing the number of layers in the circuit while keeping the rest of the algorithm unchanged (i.e., running optimization for two epochs) also fails to improve performance, likely due to the optimizer struggling with the larger parameter space. We find that a layer-wise optimization strategy, in which layers are added incrementally and only the parameters of the newly added layer are updated while previous layers are held fixed, does improve convergence. This approach reduces the effective number of parameters optimized at each step, easing the burden on the optimizer.
 
It is possible that the observed improvements from adding more layers stem primarily from the increased number of optimization steps, rather than from greater circuit expressivity. Indeed, further experiments show that much of this extra performance could be obtained by just running the algorithm to convergence with a fixed number of layers (when continuing from the best parameter configuration encountered during each epoch). However, these experiments do still suggest that additional layers might yield some improvement.

Overall, we cannot draw any strong conclusions regarding possible solutions to the decreasing algorithm performance with increasing qubit number. Using tensor network simulations, we found several promising directions, but these will need to be studied more rigorously in future work if we are to meaningfully improve the algorithm performance on larger problem instances.

Finally, we remark that the effects of hardware noise on the performance of the algorithm have not been considered in our simulations. It is likely that noise worsens the algorithm performance, but it is not clear exactly how and whether this effect is greater for larger numbers of qubits. The tensor network methods that we have described in this section are capable of simulating the effects of noise, and we leave it to future work to investigate the performance of the algorithm using noisy simulations.

\begin{figure*}
 \begin{subfigure}[b]{.25\linewidth}
    \includegraphics[width=\linewidth]{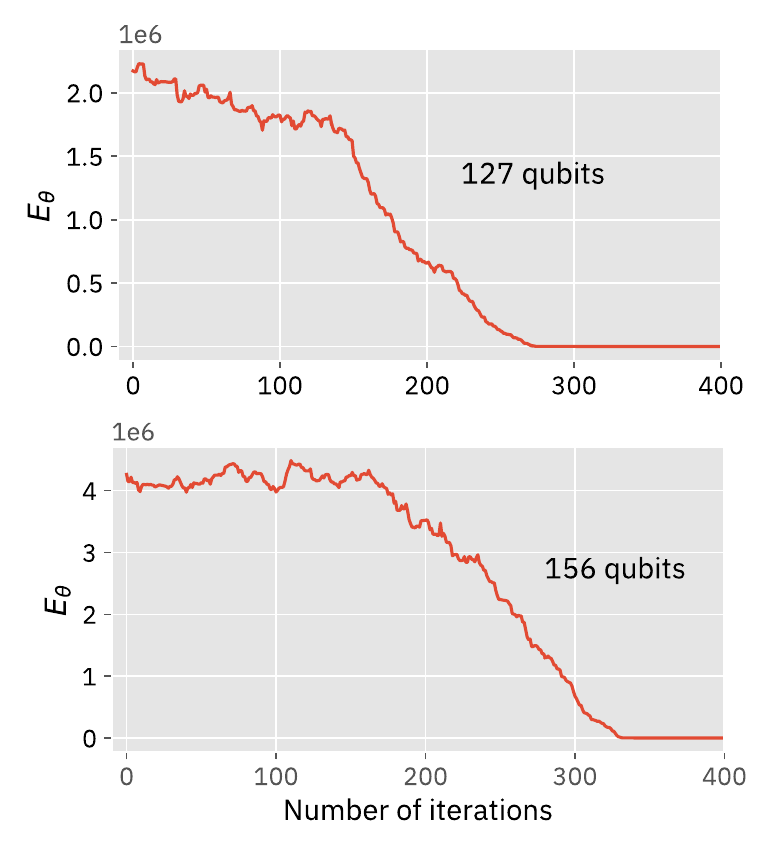}%
  \caption{}
  \label{fig: iqp evol}
  \end{subfigure}\hfill                 
\begin{subfigure}[b]{0.35\linewidth}
  \includegraphics[width=\linewidth]{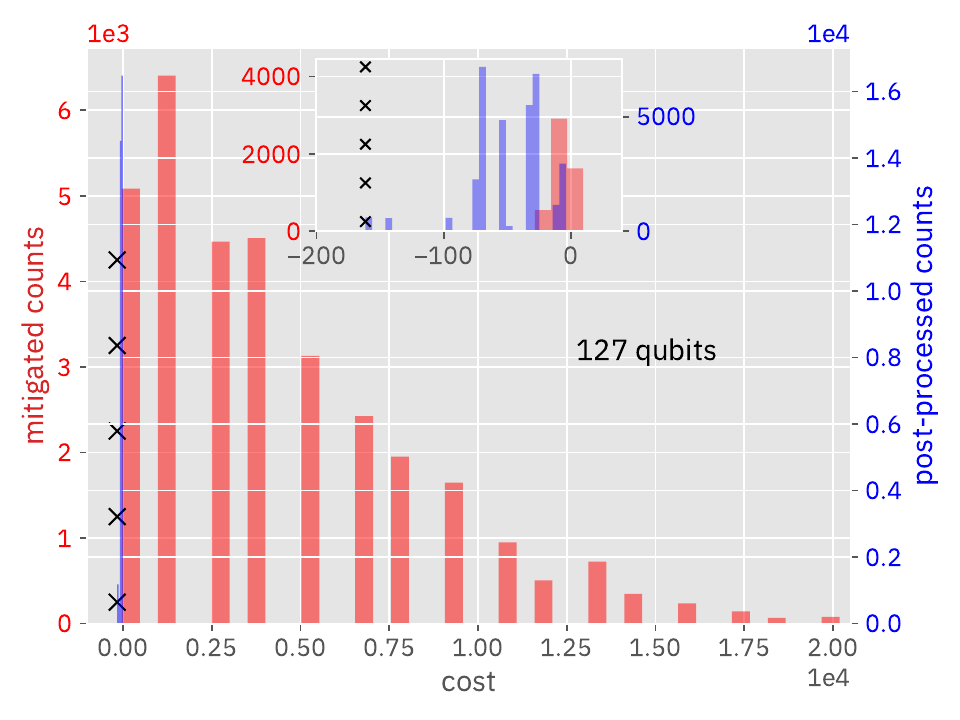}
  \caption{}
  \label{fig: iqp post n_127}
\end{subfigure}\hfill
\begin{subfigure}[b]{0.35\linewidth}
    \includegraphics[width=\linewidth]{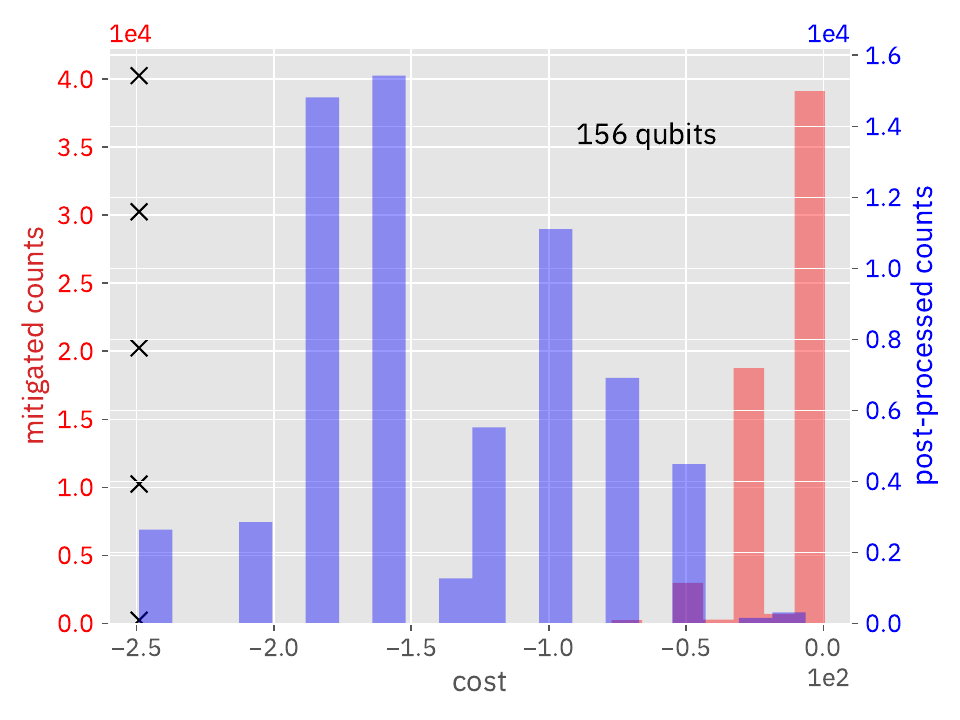}
    \caption{}
    \label{fig: iqp post n_156}
\end{subfigure}
\caption{a) Convergence plots for the expectation value $E_{\theta}$ obtained by training the variational circuit on classical nodes for the 127 and 156 qubit problem, b) and c) indicate the mitigated and post-processed (using local solver) unnormalized objective value distributions using red and blue bars, respectively. The plots correspond to hardware runs for 127 qubit and 156 qubit on \texttt{ibm\_kyiv} and \texttt{ibm\_marrakesh}, respectively. The black `x' denotes the optimal objective values of -161.5 and -249.1 for the 127 and 150 qubit problems respectively, as determined by CPLEX.}
\label{fig:raw and post-processed samples}
\end{figure*}

\subsection{IQP circuit-based optimization}
The sampling-based scheme described in Section~\ref{sampling-based} requires multiple circuit executions over several iterations on the hardware. The IQP-circuit-based optimization we propose here attempts to alleviate this problem. IQP circuits are a special class of quantum circuits characterized by commuting gates. The parameterized IQP circuit, as shown in Fig.~\ref{fig: iqp ansatz}, can be represented by a unitary $U(\theta)=H^{\otimes n} D(\theta) H^{\otimes n}$, where $D(\theta) = \prod_j e^{-i\theta_j\otimes_{i=1}^n Z_i^{m_i}}$ is a diagonal operator (in the computational basis) and bitstring $m \in \{0,1\}^n$ specifies the gate locations. $H$ and $Z$ denote the Hadamard and Pauli-Z gates, respectively.

The output distribution of an IQP circuit is generally considered classically intractable to sample from, making it a potential candidate for demonstrating quantum advantage~\cite{bremner2011classical,bremner2016average,marshall2024improved}. Interestingly, while sampling from these circuits is challenging, expectation values can be efficiently computed classically up to $poly(n^{-1})$ error~\cite{nest2009simulating}. This unique characteristic of the IQP circuit is leveraged in our approach to optimize the balance between quantum and classical computational capabilities, aiming to enhance the overall efficiency and feasibility of computational tasks.

The optimization of the parameterized IQP circuit is achieved through a variational approach, where the expectation value $E_{\theta} = \expec{0}{U^{\dagger}(\theta)H_PU(\theta)}{0}$ serves as the objective function. Given that the expectation values of observables related to our problem Hamiltonian can be efficiently computed on a classical computer, we determine the optimal circuit parameters entirely via classical computational means. Subsequently, the presumably classically challenging task of sampling is performed on quantum hardware using the circuit configured with these optimal parameters. In situations where $E_{\theta}$ does not converge sufficiently on the classical computer, one can also envision an alternate scheme where the search for better circuit parameters can be carried out using a CVaR-based sampling (see Section~\ref{sampling-based}) on the quantum computer. The ability to calculate the ideal single-qubit expectation values efficiently on a classical computer also facilitates error mitigation. This dual approach harnesses the strengths of both classical and quantum computing to improve accuracy and reliability in quantum experiments.

In Fig.~\ref{fig: iqp schematic}, we show the workflow for the quantum-centric IQP-circuit-based scheme. The task of finding the optimal parameters that minimize the expectation value $\theta_{opt} = \argmin_{\theta} \braa{\psi(\theta)}H_p\kett{\psi(\theta)}$ is carried out entirely on the classical compute nodes. Parameterized ansatz uses the diagonal operator characterized by single qubit $e^{-i\theta Z}$ and two-qubit $e^{-i\theta ZZ}$ rotations. The two qubit gates are arranged over a tubular heavy-hex layout as shown in Fig.~\ref{fig: iqp layout}. We perform $N_{iter}$ iterations of the variational scheme. Circuit parameters are initialized using random values drawn from a uniform distribution over the interval $[0, 2\pi]$. The expectation calculation for a quantum state prepared using an IQP circuit consists of computing averages over bit strings generated randomly from a uniform distribution (see Appendix~\ref{app: iqp expec}). We use a recently developed JAX-based library to compute the expectation values~\cite{recio2025iqpopt} using $2^{15}$ sample bitstrings. The circuit parameters are updated at each iteration using the NFT optimizer~\cite{nakanishi2020sequential}. The value of $N_{iter}$ is set to a sufficiently large value where the expectation values reach a plateau.

After map and optimize steps on classical nodes as outlined in Section~\ref{sampling-based}, the circuit parameterized by $\theta_{opt}$ is executed on the quantum processor. The raw samples thus obtained are then subjected to post-processing on classical nodes. Initially, this involves using an IQP-circuit-based approach for error mitigation. Given that single-qubit observables from IQP circuits can also be computed efficiently, we use the classically computed noiseless single-qubit expectation values $\langle Z_i \rangle$ to correct the samples. For qubits where the absolute expectation values 
$\abs{\langle Z_i \rangle}$ exceed a predefined positive threshold $\langle Z_i \rangle_{th}$, which we set at 0.99, the qubit is highly likely to be in a state of $\kett{0}$ or $\kett{1}$. Such qubits are accordingly set to 0 or 1 in the raw samples, while those not meeting this threshold are retained as is. In the subsequent post-processing phase, the mitigated samples undergo further refinement through a local search technique as described in Section~\ref{sec: post processing}. 

\subsubsection{Hardware results}
In order to examine the performance of the proposed scheme we performed experiments on 127 and 156 qubit problems corresponding to mRNA sequence lengths of 45 and 60 nucleotides, respectively. We use the reduction scheme (Appendix~\ref{app: reduction}) to create the 156 qubit problem from an original 256 qubit problem. The 127 and 156 qubit experiments were conducted on IBM's Eagle processor \texttt{ibm\_kyiv} and Heron r2 processor \texttt{ibm\_marrakesh}, respectively. These circuits are characterized by total circuit depths up to 112 and consists of approximately 450 non-local gates. We use dynamical decoupling with the default "XX" sequence for error suppression. In Fig.~\ref{fig: iqp evol}, we show the evolution of the expectation values $E_{\theta}$ during the variational scheme carried out on the classical nodes. As observed in the plots, $E_{\theta}$ gradually reaches a plateau after approximately 320 iterations. In Figs.~\ref{fig: iqp post n_127} and~\ref{fig: iqp post n_156}, we present the mitigated and post-processed unnormalized objective value distributions for the 127 and 156 qubit experiments. It is clear from the plot that the mitigated hardware distribution is already relatively close to the ground state. This allows the local search scheme to find the ground state with a relatively high probability.

In contrast with the line-based connectivity utilized for the two-local ansatz (refer to Section~\ref{sampling-based}), we employ a tubular heavy-hex structure here. Deviation from the one-dimensional circuit layout serves as the first step to discover circuits with better convergence properties that are at the same time also hard to classically simulate for sampling tasks. 

Our results provide evidence that parameterized IQP circuit-based optimization can serve as a promising tool to investigate optimization problems.

\section{Conclusions and outlook}\label{sec:conclusions}

In this study, we have introduced a quantum-centric optimization workflow that strategically utilizes both quantum and classical computing resources to effectively tackle the problem of predicting the secondary structure of longer mRNA sequences. Our approach, which incorporates a novel IQP-based sampling scheme, not only facilitates good convergence but also takes a step further towards delegating classically difficult tasks to the quantum computer. Additionally, the custom error mitigation strategies we developed significantly enhance the quality of the noisy samples obtained from quantum hardware, motivating further exploration of such techniques in sampling-based approaches and other problem domains.

Tensor network simulations of the CVaR VQA played a supportive role in this study, contributing to the validation of the quantum algorithm performance. By enabling simulations of larger quantum systems, tensor networks have complemented the capabilities of current quantum hardware, especially in scenarios involving an increased number of qubits. This support has helped to evaluate the effectiveness of these approaches and identify areas for potential methodological improvements.

Our work indicates that problem-agnostic circuits based on the hardware topology are of practical relevance. However, the theoretical foundations of such fixed-architecture circuits remain less explored~\cite{farhi2017quantum}. 
Future research will aim to strategically design connectivity layouts that enhance performance.

Our implementation demonstrates that the variational scheme can reach sufficient convergence, leading to circuits that can create samples in the vicinity of the ground state. However, due to the heuristic nature of such schemes, it is difficult to provide performance guarantees as the problems scale. This underscores the need for more extensive testing with larger problem instances. Additionally, we plan to explore non-variational schemes such as Bias Field Counterdiabatic scheme~\cite{cadavid2024bias,romero2024bias} and quantum enhanced MCMC~\cite{layden2023quantum}, which rely on problem-inspired circuits. These methods necessitate the development of advanced techniques for embedding these problem-specific circuits into the existing quantum hardware topology, aiming to minimize the overhead associated with gate counts.

\section*{Acknowledgments}
The authors thank Yashrajsinh Jadeja and Haining Lin for their contributions to the prediction of classical RNA secondary structures. Fermioniq was supported by the Dutch National Growth Fund (NGF), as part of the Quantum Delta NL Programme.

\appendix
\subsection{Reduction Scheme}
\label{app: reduction}

We describe a problem reduction scheme we have developed to reduce the number of problem variables. Let the mRNA secondary structure prediction problem, as described in Section~\ref{sec:QUBO}, has its general QUBO form, 
$Q = \sum_i h_i x_i + \sum_i \sum_{j(i)} J_{ij}x_i x_j.$

The proposed reduced scheme is possible because of the following two properties of $Q$.
\begin{enumerate}
    \item The coefficients of the linear terms are much smaller in magnitude than the coefficients of the quadratic terms i.e. $\abs{h_i} \ll \abs{J_{ij}}$
    \item $Q$ contains a subset $\kappa$ of variables $k$ whose quadratic coefficients $J_{kj} \ge 0$ for $\forall~ j(k)$.
\end{enumerate}
Due to the properties above, for any assignment of variables in $\in \kappa$, their contributions to $Q$ will always be $\ge 0$. As a result, one can only focus on variables $\notin \kappa$ and formulate a reduced QUBO, $Q_{\mathrm{red}} = \sum_{i \notin \kappa}\big(h_i x_i + \sum_{j(i) \notin \kappa} J_{ij} x_i x_j\big)$
containing $n - \abs{\kappa}$ variables where $n$ is the original problem size. After we have found the optimal solution to $Q_{\mathrm{red}}$, the task at hand is to find the assignment of variables $\in \kappa$ that gives the lowest cost. This in turn requires finding the optimal solution to

\begin{align}
\label{eq:chi part}
Q_{\kappa} = \sum_{i \in \kappa}\big(h_i x_i + \sum_{j(i) \in \kappa} J_{ij} x_i x_j + \sum_{j(i) \notin \kappa} J_{ij} x_i x_j \big)
\end{align}

As we have already found the optimal assignments of the variables $\notin \kappa$ by solving $Q_{\mathrm{red}}$, this information can be used to find the optimal assignment of the variables $\in \kappa$. It is clear from Eq.~(\ref{eq:chi part}), if $x_j = 1$ in $\sum_{j(i) \notin \kappa} J_{ij} x_i x_j$, then $x_i$ must be assigned a value of $0$ because setting them to $1$ is guaranteed to increase the objective $Q_{\kappa}$. Only the subset of variables $\kappa$ whose quadratic contribution $\sum_{i \in \kappa} \sum_{j(i) \notin \kappa} J_{ij} x_i x_j$ is equal to zero (owing to $x_j$s equal to zero) and the coefficient of the linear term $h_i < 0$ can lower the value of $Q_{\kappa}$. In order to find their optimal assignment, one needs to solve another QUBO, with the objective function as$Q_{\kappa} = \sum_{i \in \kappa}\big(h_i x_i + \sum_{j(i) \in \kappa} J_{ij} x_i x_j\big).$ 
 In our problems, the number of variables $\in \kappa$ is relatively small and can usually be solved by exact enumeration.
 
\subsection{Computing expectations from parameterized IQP circuit}
\label{app: iqp expec}
If the IQP circuit is denoted by the unitary $U(\theta)=H^{\otimes n}\prod_j e^{i\theta_j\otimes_{i=1}^n Z_i^{m_i}} H^{\otimes n}$, where $m \in \{0,1\}^n$, the expectation $E_{\theta}$ corresponding to the observable $\bigotimes_{k=1}^n Z_k^{p_k}$, where $p_k \in \{0,1\}^n$ can be written as

\begin{align}
    &\braa{+}^{\otimes n} \!\prod_j\! e^{i\theta_j\otimes_{i=1}^n Z_i^{m_i}} \!\!H^{\otimes n}\bigotimes_{k=1}^n Z_k^{p_k}H^{\otimes n} \!\prod_{j}\! e^{-i\theta_{j}\otimes_{i=1}^n Z_i^{m_i}} \kett{+}^{\otimes n} \nonumber \\
    &= \frac{1}{2^n}\!\!\!\!\sum_{x \in \{0,1\}^n}\!\!\!\!e^{i \sum_j \theta_j (-1)^{m \cdot x}}\!\braa{x}\bigotimes_{k=1}^n X_k^{p_k}\!\!\!\!\sum_{y \in \{0,1\}^n}\!\!\!\!e^{-i \sum_j \theta_j (-1)^{m \cdot y}}\!\kett{y} \nonumber \\
    &= \frac{1}{2^n}\!\!\!\sum_{x \in \{0,1\}^n}\!\!\!e^{i \sum_j \theta_j (-1)^{m \cdot x}}\braa{x}\!\!\!\sum_{y \in \{0,1\}^n}\!\!\!e^{-i \sum_j \theta_j (-1)^{m \cdot y}}\kett{y\oplus p} \nonumber \\
    &= \frac{1}{2^n}\!\!\!\sum_{y \in \{0,1\}^n}\!\!\!e^{i \sum_j \theta_j (-1)^{\sum_k m_k(y_k\oplus p_k-y_k)}} 
\end{align}
Note $(\cdot)$ denotes bitwise dotproduct and $\oplus$ denotes addition modulo 2.
As expectation is guaranteed to be real valued, we can write 

\begin{align}
   E_{\theta} = \frac{1}{2^n}\sum_{y \in \{0,1\}^n} \cos{\bigl[\sum_j \theta_j (-1)^{\sum_k m_k(y_k\oplus p_k-y_k)}\bigr]},
\end{align}

which can be calculated by computing the empirical mean $\mathbb{E}_{y \sim U} \cos{\bigl[\sum_j \theta_j (-1)^{\sum_k m_k(y_k\oplus p_k-y_k)}}\bigr]$ over the sample bitstrings $y$ drawn from a uniform distribution $U$.

\bibliographystyle{myIEEEtran}
\bibliography{ref}

\end{document}